\documentclass[aps, twocolumn]{revtex4-1}

\usepackage{graphicx}
\usepackage{amsmath}
\usepackage{dsfont}
\usepackage{amssymb}
\usepackage{physics}
\usepackage{hyperref}
\usepackage{ulem}
\hypersetup{colorlinks=true,
	    final=true,
	    linkcolor=blue,
	    citecolor=blue,
	    filecolor=blue,
	    urlcolor=blue,}

\usepackage{amssymb}

\newcommand{\ep}{\epsilon}

\newcommand{\p}{\prime}

\newcommand{\ra}{\rangle}
\newcommand{\la}{\langle}

\newcommand{\beq}{\begin{equation}}
\newcommand{\eeq}{\end{equation}}

\newcommand{\mck}{\mathcal{K}}

\newcommand{\upa}{\uparrow}

\newcommand{\downa}{\downarrow}

\begin{document}

\title{V. J. Emery and P. W. Anderson's views and related issues regarding the basics of cuprates: a re-look}
\author{Navinder Singh}
\email{navinder.phy@gmail.com; Phone: +91 9662680605}
\affiliation{Theoretical Physics Division, Physical Research Laboratory (PRL), Ahmedabad, India. PIN: 380009.}

\begin{abstract}
In 1991, V. J. Emery in his important review article entitled "Some aspects of the theory of high temperature superconductors"\cite{emery1} argued against the Zhang-Rice reduction of three-band to an effective one-band model. In his words "...therefore it seems that the simple $t-J$ model  does not account for the properties of high temperature superconductors". Over approximately 35 years after the initial debates\cite{debates} much has happened in the field pertaining to this topic. Even though it is one of the most discussed issue, a comprehensive account and the required resolution are lacking. Connected to the debate over one-band versus three-band models is another discussion: the one-component versus two-component model for cuprates. The two-component model is most strongly advocated by Barzykin and Pines\cite{bp}.  In this article the author attempts a perspective and a re-look on some of these issues. After an analysis of a large body of literature, author finds that V. J. Emery's criticism of the Zhang-Rice reduction was correct. Many central experimental features of cuprates cannot be rationalized within the one-band model, and Johnston-Nakano scaling is one such example. Other examples are also discussed. Author introduces a simple-minded toy model to illustrate the core issues involved.
\end{abstract}

\maketitle

\begin{widetext}
\begin{quote}
{\texttt{``No theory has been able to explain why most of the doped or metallized antiferromagnetic insulators such as $LaTiO_3, ~V_2O_3,~NiS_2,$ and $Sr_2VO_4$ do not show superconductivity even at very low temperatures. Neither the relatively large antiferromagnetic superexchange interaction nor strong two-dimensionality has been able to explain such a dramatic difference convincingly.''   --- Masatoshi Imada, Nat. Phys. Vol. 2, 138 (2006). } }
\end{quote}
 \end{widetext}

 \section{Introduction} 

 To set the stage let us first very briefly review the basic electronic structure of cuprates. One of the most clear and rigorous account of it is presented in\cite{mott}. We call this the standard paradigm. The argument for $La_2CuO_4$ goes in the following manner: 
 
 In $La_2CuO_4$ each $Cu$ ion is surrounded by 6 oxygen ions (out of which 4 oxygen ions are in the $CuO_2$ plane, and two apical oxygen ions are above and below the $CuO_2$ plane, thus forming an octahedron). The perfect octahedron  is distorted in such a way that the apical oxygen atoms are pushed away from the $CuO_2$ plane and 4 planner oxygen ions are pulled-in towards the copper ion.  The reason is the net lowering of energy (the Jahn-Teller distortion) and this distortion lifts the degeneracy of the top most copper orbitals $3d_{x^2-y^2}$ and $3d_{3z^2-r^2}$ in such a way that highest energy orbital is $3d_{x^2-y^2}$ which contains a single electron or a single hole. 
 
 \begin{figure}[h!]
    \centering
    \includegraphics[width=1.0\columnwidth]{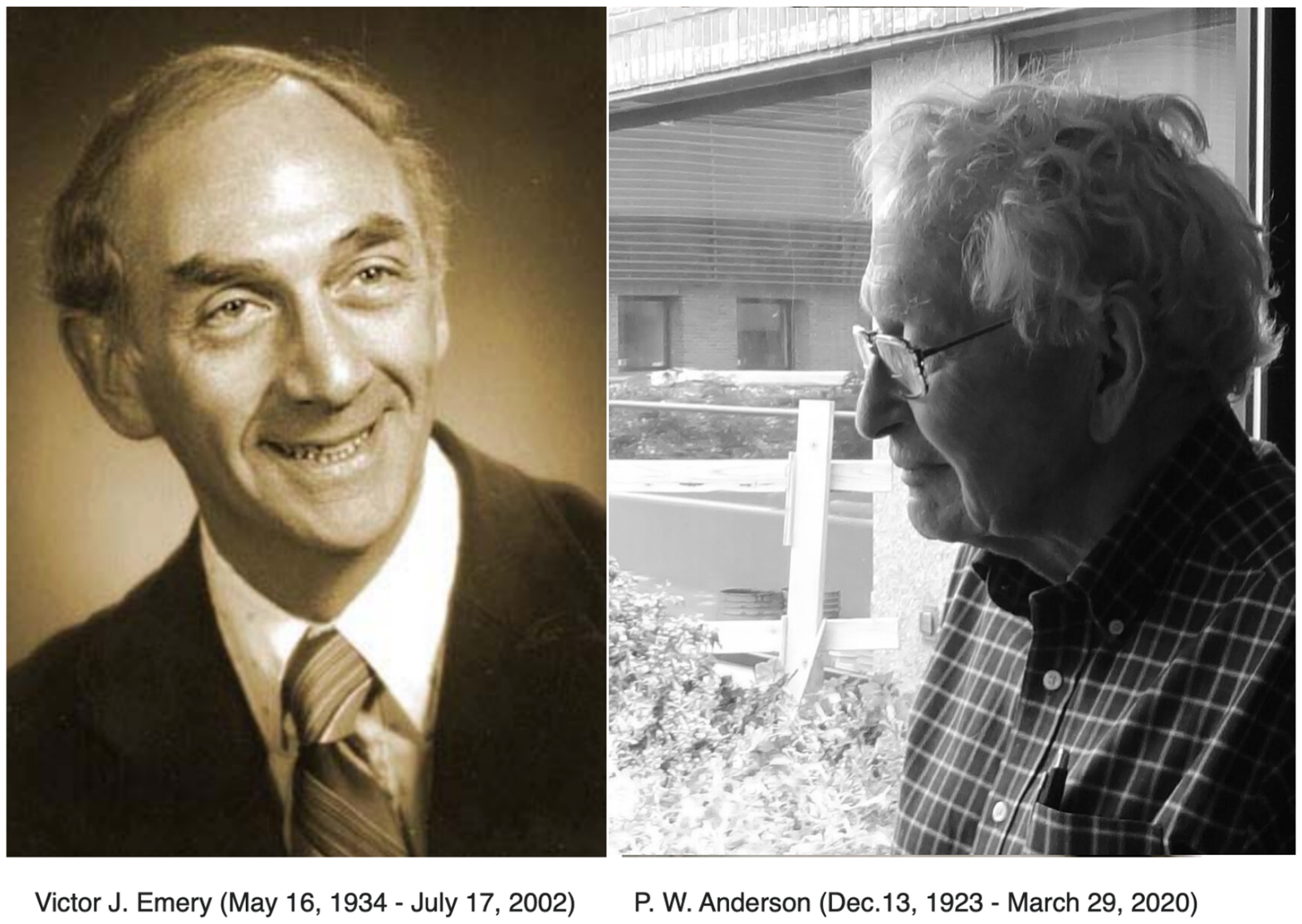}
    \caption{Through this article, the author respectfully honors two pioneering figures in the field for their invaluable contributions. Photo Courtesy: V. J. Emery (from John Tranquada with thanks) and P. W. Anderson (from G. Baskaran with thanks).}
    \label{f1}
\end{figure} 
 
 Lobes of this $3d_{x^2-y^2}$ copper orbital point towards lobes of $p_x$ and $p_y$ orbitals of planner oxygen ions leading to strong hybridization between them. Consequently a large hopping integral $t_{pd}$ exists. In the hole picture (figure 2) the single hole when situated in copper $3d_{x^2-y^2}$ orbital has energy $\ep_d$ (say) and when it migrates to oxygen p-orbitals it has energy $\ep_p,~~(\ep_p>\ep_d)$. If the hopping integral $t_{pd}$ is greater than $\Delta_{CT} = \ep_p -\ep_d$, then the hole can migrate to oxygen orbitals. In other words, the strong hybridization leading to $t_{pd}$ can delocalize the single hole on copper $3d_{x^2-y^2}$ orbital if the above condition is met. However, it turns out that $\Delta_{CT}$ (called the charge transfer energy) is greater than the hopping integral $t_{pd}$ and therefore the hole stays localized on copper. Thus,  the undoped system will have a localized holes at copper sites with spin $S= \frac{1}{2}$ and oxygen orbitals are completely filled. In the undoped state localized spins on the copper atoms form an AFM lattice (via superexchange mechanism). 
\begin{figure}[h!]
    \centering
    \includegraphics[width=1.0\columnwidth]{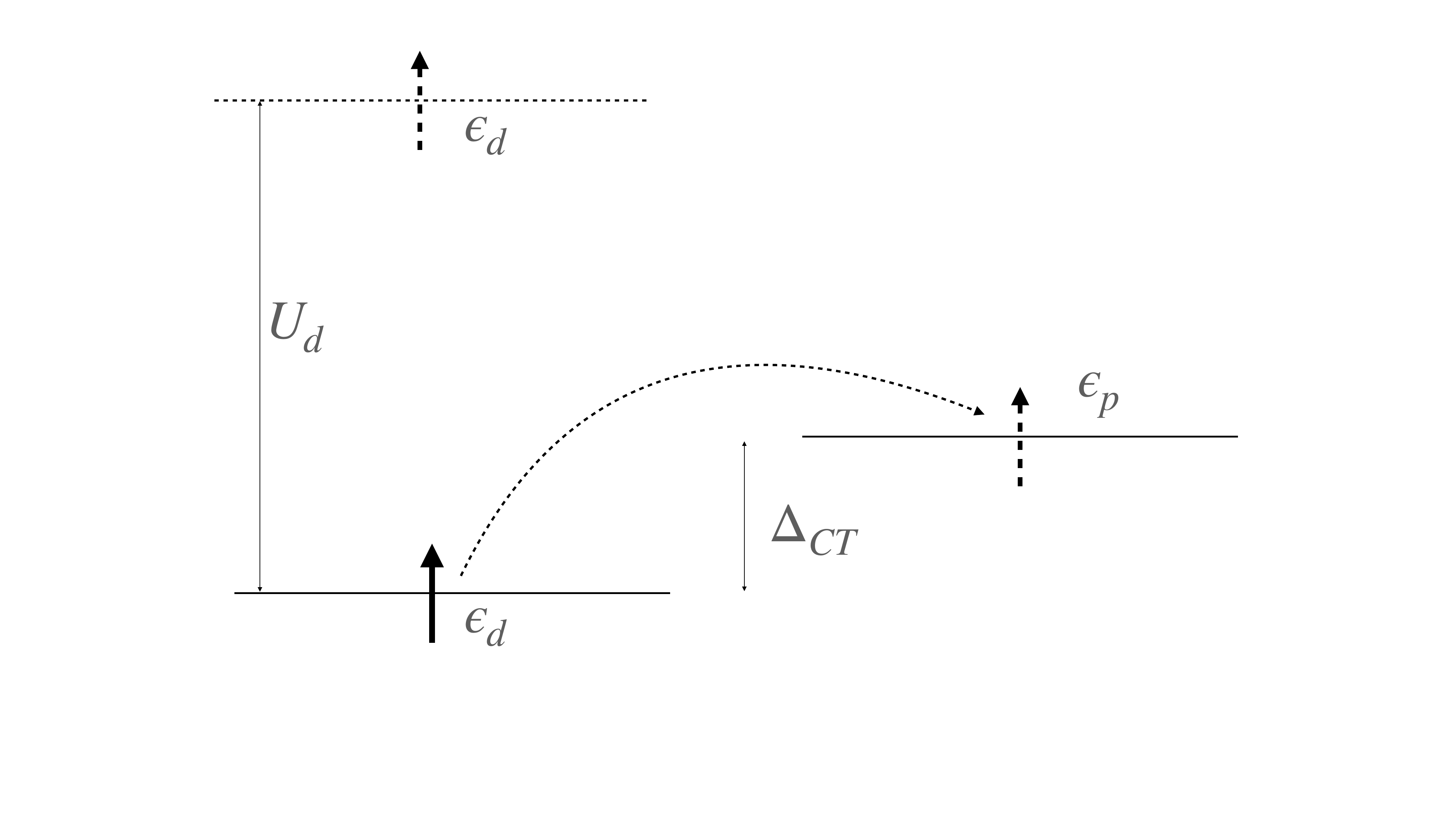}
    \caption{A schematic energy level diagram.}
    \label{f2}
\end{figure}

The question is what will happen when hole doping is done (i.e., when a fraction of electrons removed from $CuO_2$ layers)? Here these two pioneers (P. W. Anderson and V. J. Emery) differ:

\section{P. W. Anderson's point of view}

Anderson maintained\cite{ander1} that the highest energy band is formed by anti-bonding $Cu 3d_{x^2-y^2} --O2p_\sigma$ hybrid orbitals, and the Fermi energy passes through it. At zero doping copper holes are localized due to strong onsite Hubbard interaction, and superexchange via ligand oxygen ions leads to AFM ordering of spins of the localized copper holes.  Hole doping removes the lone electrons in copper $3d_{x^2-y^2}$ orbitals making them doubly occupied by holes ($Cu^{+++}$, one missing from $4s$ orbital and two from the $3d$ orbital). The doped hole is a mobile object (flits from one copper site to another). In his own words\cite{ander1,anderbook}:

\begin{quote}
{\texttt{``...then what seems likely is that some of the $Cu$ will become $Cu^{+++}$, a possible, if not a very common, valence for copper. Right from the start, I assumed that it was these positively charged "holes" in the $Cu^{++}$ d-shell which were the metallic carrier -- as had the inventors, but not everyone else.''   --- P. W. Anderson\cite{ander1} } }
\end{quote}

Anderson further argues that the holes in the $Cu^{++}$ d-shells act both as magnetic degree-of-freedom at zero and at very low doping, and as mobile degree-of-freedom at finite doping (on doping, $Cu^{++}$ goes to $Cu^{+++}$). That is, these are responsible for conductivity and superconductivity at a finite doping. He calls it "either-or" basis\cite{anderbook,comment1}:

\begin{quote}
{\texttt{``...superconductivity is exhibited by the same structures, and literally the same electrons, which show simple magnetic ordering, on what must quite clearly be seen as an "either-or" basis: either magnetic and insulating, or superconducting and metallic,...."    --- P. W. Anderson\cite{anderbook, comment1}.
}}
\end{quote}

The model thus advocated by Anderson\cite{ander1} is the single band Hubbard model:

\beq
H + T + U \sum_i n_{i\uparrow} n_{i,\downarrow}.
\eeq
Constituted by the anti-bonding $Cu 3d_{x^2-y^2} --O2p_\sigma$ hybrid orbitals with narrow bandwidth and having strong electronic correlation effects in it. In the above equation, $T = t\sum_{<i,j>} d_{i\sigma}^\dagger d_{j\sigma} +h.c.$, is the kinetic energy term, and the second term is the on-site Hubbard interaction term. In the large $U$ limit, this Hamiltonian can be canonically transformed to the $t-J$ Hamiltonian\cite{anderbook}: 

\beq
H_{t-J} = P T P + J \sum_{i,j} S_i.S_j,
\eeq
in which double occupancy is projected out by using the Gutzwiller projection operator: $P = \Pi_i (1- n_{i\upa}n_{i\downa})$. We will observe that some of the very fundamental experimental features cannot be rationalized within the one-band model (section(V)).

\section{V. J. Emery's point of view}

Right from the start,  V. J. Emery\cite{emery2} (also Varma etal\cite{varma1}) assumed that doped holes go into oxygen $2p$ orbitals (rather than in the copper $3d_{x^2-y^2}$ orbital) and these constitute itinerant degrees-of-freedom. The copper holes (holes in $3d_{x^2-y^2}$ orbitals localized due to $\Delta_{CT}$)  stay localized and constitute AFM background. Due to $d-p$ hybridization, mobile $p$ holes pick up some $d$ character. But these doped holes remain mobile in the "empty" oxygen sub-lattice ($\grave{a} ~la~ $Emery\cite{emery1}).

Therefore, Emery\cite{emery1} and Varma etal\cite{varma1} argued that proper description of cuprates require the explicit inclusion of oxygen $2p_x,~2p_y$ orbitals in addition to the $Cu 3d_{x^2-y^2}$ orbitals. These considerations lead to the celebrated three-band model:

\begin{eqnarray}
H &=&  \sum_{i,j,\sigma} \ep_{ij} a_{i,\sigma}^\dagger a_{j,\sigma} + \frac{1}{2} \sum_{i,j,\sigma,\sigma^\p} U_{i,j} a_{i,\sigma}^\dagger a_{i,\sigma} a_{j,\sigma^\p}^\dagger a_{j,\sigma^\p}. \nonumber\\
 &=& \sum_{j,\sigma}\ep_p p_{j,\sigma}^\dagger p_{j,\sigma} + \sum_{i,\sigma}\ep_d d^\dagger_{i,\sigma}d_{i,\sigma}\nonumber\\
 &+& U_p\sum_{j,\sigma,\sigma^\p} p_{j,\sigma}^\dagger p_{j,\sigma} p_{j,\sigma^\p}^\dagger p_{j,\sigma^\p}+U_d\sum_{i,\sigma,\sigma^\p} d_{i,\sigma}^\dagger d_{i,\sigma} d_{i,\sigma^\p}^\dagger d_{i,\sigma^\p}\nonumber\\ 
 &-&t_{pd} \sum_{i,j,\sigma}(d_{i,\sigma}^\dagger p_{j,\sigma} + h.c.) +\frac{1}{2} V \sum_{i,j,\sigma,\sigma^\p} p_{j,\sigma}^\dagger p_{j,\sigma}  d_{i,\sigma^\p}^\dagger d_{i,\sigma^\p}.\nonumber\\ 
 \end{eqnarray}

 Vacuum state is chosen such that all $d$ states occupied ($Cu^+$) and all $p$ states occupied $O^{2-}$. Creation and annihilation operators are for holes. First term in the second line in the above equation array denotes site energies for holes on oxygen ($\ep_p$), and second term in second line denotes site energies for holes on copper ($\ep_d$). $U_p$ and $U_d$ in the third line denote on-site Hubbard interactions at oxygen sites and at copper sites, respectively. First term in the last line denotes the overlap integral due $d-p$ hybridization, and last term is the Coulomb repulsion between a hole on copper and another hole on nearby oxygen. The index $i$ is for copper sub-lattice, and the index $j$ is for the oxygen sub-lattice.

From a different line of arguments rooted in the experiments\cite{bp}, it has been observed that $CuO_2$ planes show a behaviour that can be mapped to what is called "two-component" model. Barzykin and Pines\cite{bp}, based on a detail analysis of a variety of experimental probes, noticed that $CuO_2$ planes exhibit the behaviour that can be divided into two very different characters: (1) Fermi-liquid like component, and (2) spin liquid component.  In section (V) we will observe that this three-band Emery model naturally leads to the two-component description at the phenomenological level which Barzykin and Pines stressed.

Next, In 1988, Zhang and Rice argued that the doped hole in oxygen $p-$orbitals will resonate with already existing unpaired hole on copper $d$ orbital. It will have singlet and triplet states. The singlet state (ZR singlet) has the lower energy,  and this "spin zero composite object" will hop from one copper-oxygen plaquette to another, with near neighbor hopping $t$ and next near hopping $t^{\p}$ etc. The reduced Hamiltonian (from three-band to one-band) can be written as in equation (2), the $t-J$ model. The important point to be noted is that in Zhang-Rice picture doping holes remove spins (singlets have zero spin) from the $CuO_2$ plane whereas in the Emery model hole doping adds extra spins. Zhang-Rice reduction has been controversial\cite{debates}. In subsequent sections we will analyze an array of experimental results that do not support ZR reduction. However, before that, and to gain some analytic insight for the case of a doped hole into the system, it is useful to study a toy model.

\section{A toy model}

To understand why some of the experimental evidence is against the Zhang-Rice singlet picture, we consider a very simple minded toy model in which we will notice that the doped carrier keeps on "running" (without forming a local singlet) in the lattice. For the time being let us focus on the following toy model (and not compare it with the actual realistic situation in $CuO_2$ planes). We will comment on it after analyzing this toy model.

\begin{figure}[h!]
    \centering
    \includegraphics[width=1.0\columnwidth]{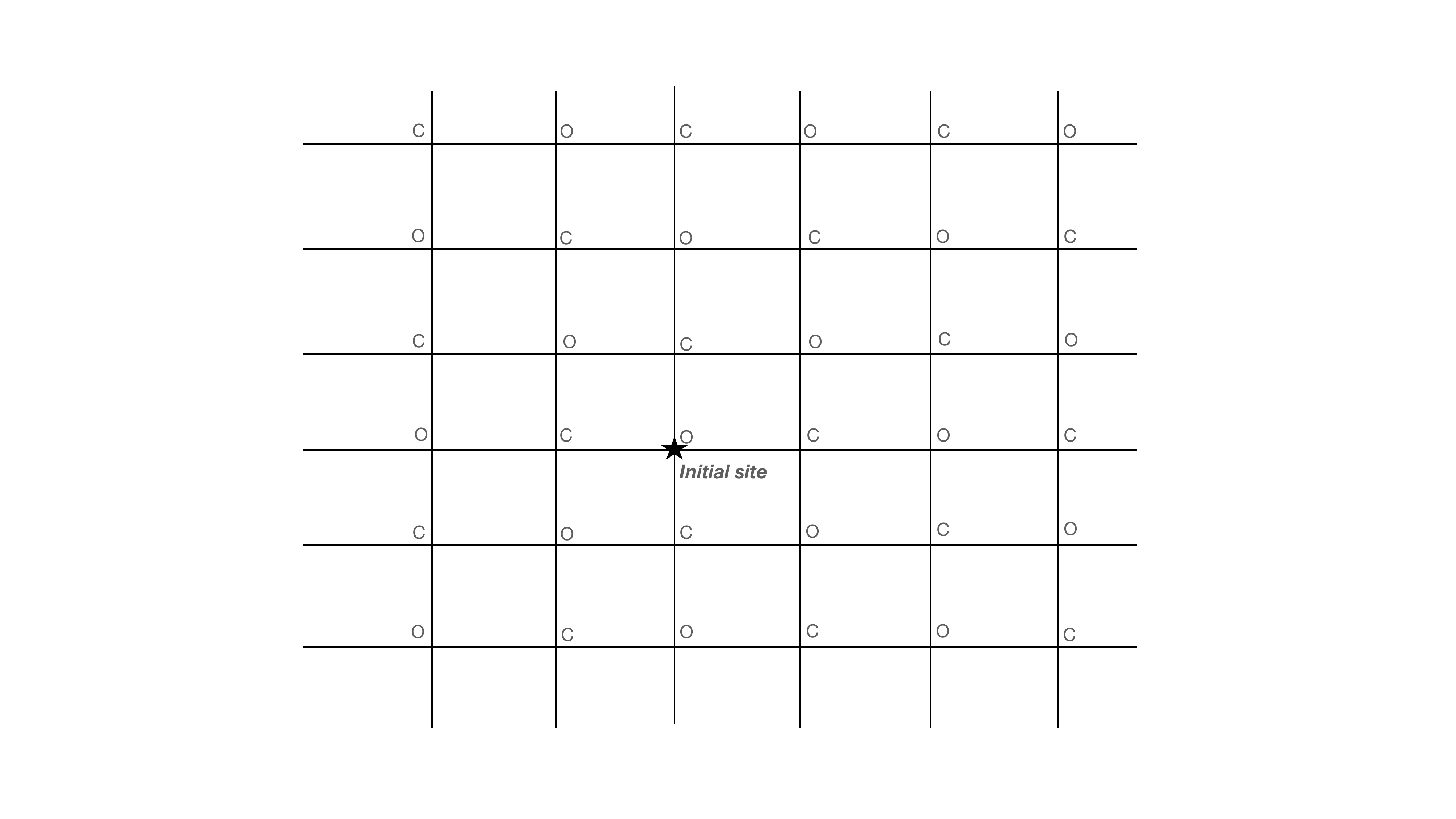}
    \caption{Schematic of the toy model.}
    \label{f3}
\end{figure}

Suppose that there is a 2D square lattice with two types of sites: C-type and O-type (refer to figure 3). Consider that at an initial time $t=0$ a particle is placed at an O-site. It can tunnel to any neighboring C-type sites with amplitude $-t$. It's onsite energy when it is situated on O-type sites is set equal to zero. But when it goes to C-type sites (due to some "unknown mechanism")  it's energy becomes $U>0$.

We would like to ask the following questions. Lets us assume that $|t|<<U$:

\begin{enumerate}
\item Will the particle stay at the initial O-type site, forever? (it is sitting in a local minima of energy!).
\item Or, will it hop (flit) to C-type site and come back to the original O-type site, and repeat this motion forever (in Linus Pauling's words, will it set into a "resonance")? Figure 4 A.
\item Or, will it hop to C-type site and rather than coming back to the original O-type, it will hop to diametrically opposite O-type site and come back again following its original path and repeat the motion (i.e., set in a resonance over three sites O-C-O ( $|1\rangle$, $|2\rangle$, and $|3\rangle$); Figure 4 B)?
\item Or, will it set into "loop" kind of resonance? Or a "star" kind of resonance (Figure 4 C and D).
\item Or, will it keep running into the lattice without forming any resonance? Or forming a very extended resonance? 
 \end{enumerate}

\begin{figure}[h!]
    \centering
    \includegraphics[width=1.0\columnwidth]{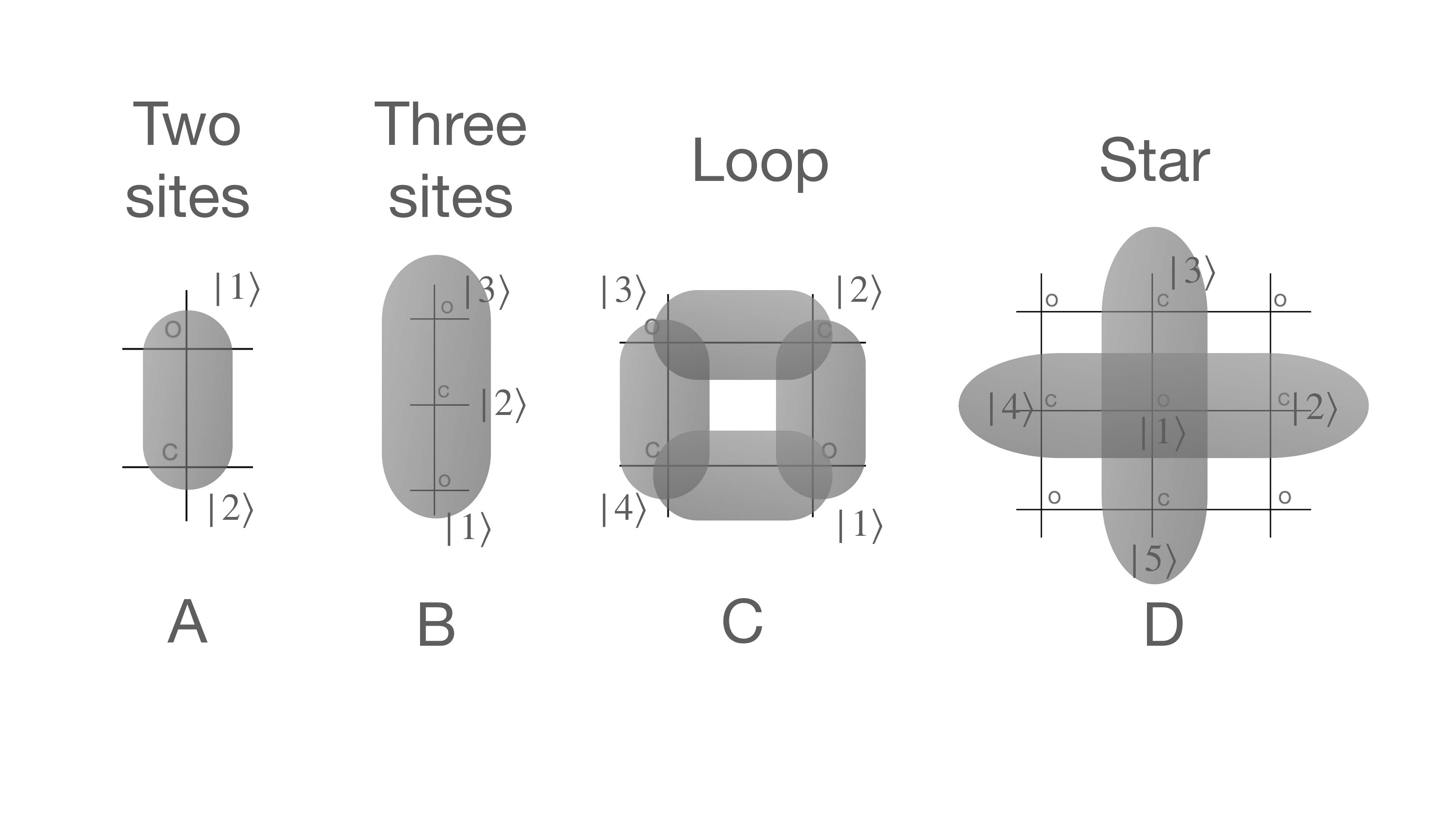}
    \caption{Some of the various possible resonances.}
    \label{f4}
\end{figure}

To analyze the problem, consider each case one by one.

\subsection*{Case A: Resonating over $|1\rangle$ and $|2\rangle$.}
Look at figure 4A. There are two states $|1\ra$ and $|2\ra$. These are base kets and the Hilbert space is 2 dimensional. The Hamiltonian can be written as $H = 0 |1\ra\la1| + U |2\ra\la2| - t (|1\ra\la2| +h.c.)$. In the representation $|1\ra = \big(\begin{smallmatrix} 1\\0\end{smallmatrix}\big)$ and $|2\ra = \big(\begin{smallmatrix} 0\\1\end{smallmatrix}\big)$, the Hamiltonian is a $2 \cross 2$ matrix with eigenvalues and eigenkets as: $\lambda_{-} = \frac{1}{2} (U - \sqrt{U^2 + 4 t^2}) [~with~|-\ra = \frac{1}{\sqrt{2}} (|1\ra - |2\ra)]$, and $\lambda_{+} = \frac{1}{2} (U + \sqrt{U^2 + 4 t^2}) [~with~|+\ra = \frac{1}{\sqrt{2}} (|1\ra + |2\ra)]$. Clearly, in the limit $|t|<<U$, the lowest eigenvalues is

\beq
\lambda_{-}  \simeq -\frac{t^2}{U}.
\eeq

Negative energy signifies a bound state. But, is it the only possible bound state? Or will some other motion of the particle in the lattice lead to even lower energy? To understand it, consider case B.

\subsection*{Case B: Resonating over $|1\rangle$, $|2\rangle$, and $|3\rangle$.}

The resonance over three states $|1\ra,|2\ra,~|3\ra$ is shown in figure 4B. The Hamiltonian in this case can be written as: $H = 0 |1\ra\la1| +U |2\ra\la2| + 0 |3\ra\la3| - t (|1\ra\la2| +h.c.) - t (|2\ra\la3| +h.c.)$. It can be esily diagonalized to get three eigenvalues:

\begin{eqnarray}
\lambda_0 &=& 0\nonumber\\
\lambda_{-} &=& \frac{1}{2} (U - \sqrt{U^2 +8t^2})\nonumber\\
\lambda_{+} &=& \frac{1}{2} (U + \sqrt{U^2 +8t^2}).
\end{eqnarray}

The lowest eigenvalue in the limit $|t|<<U$ is:

\beq
\lambda_{-} \simeq -2 \frac{t^2}{U}.
\eeq

Clearly, it is lower than that in case A. The particle, rather than settling in a singlet resonance over $|1\ra$ and $|2\ra$, will rather settle in resonance over three states  ($|1\rangle$, $|2\rangle$, and $|3\rangle$)  with lowest eigenvalue ($\lambda_{-}$).

\subsection*{Case B1: Resonating over $|1\rangle$, $|2\rangle$, and $|3\rangle$. But the state on the central C-type site is a virtual state (figure 4B)}

If the particle is in resonance over two O-type sites and the central site (C-type site) constitutes only a virtual state, then will it further lower its energy as compared to than that in Case B? Clearly the hamiltonian in this case is:
$H = 0 |1\ra\la1| + 0 |3\ra\la3| - t' (|1\ra\la3| +h.c.)$. That is, both O-type sites have zero energy and C-type state is virtual. The tunneling is now represented as $-t'$. The Hamiltonian is easily diagnosable leading to two eigenvalues: $\lambda_{-} = -t'$, and $\lambda_{+} = + t'$.

The situation here is akin to what happens in an ammonia molecule (figure 5).
\begin{figure}[h!]
    \centering
    \includegraphics[width=0.9\columnwidth]{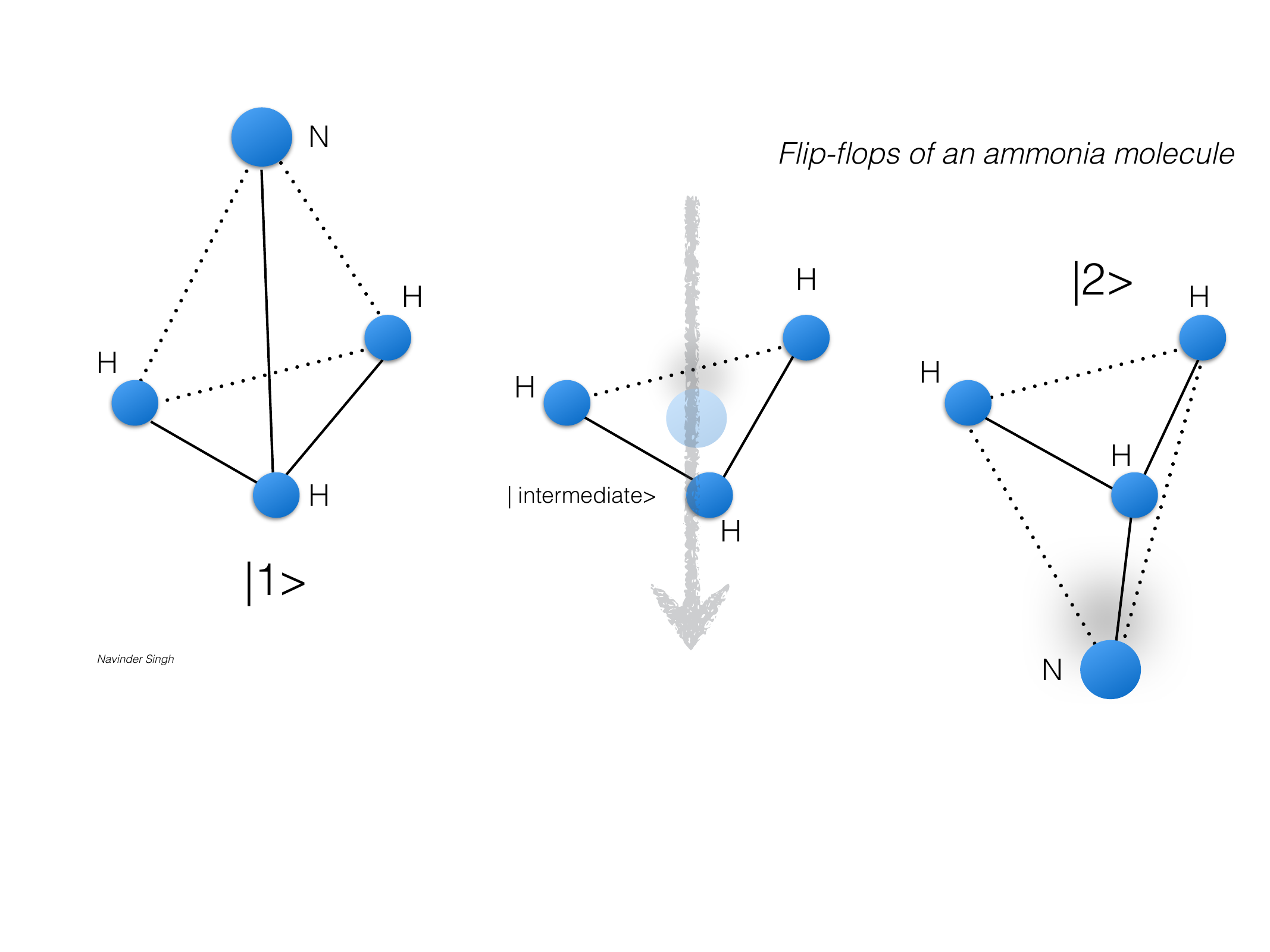}
    \caption{Ammonia molecule.}
    \label{f5}
\end{figure}
In one state of it (say, $|1\ra$), nitrogen atom is situated above the plane formed by three hydrogen atoms. What happens in an ammonia molecule is that the nitrogen atom quantum mechanically tunnels through the plane formed by the three hydrogen atoms and can appear on the other side of it constituting another state, say $|2\ra$ (figure 5). Feynman called it ``flip-flops" of the nitrogen atom. The quantum state of the molecule is the superposition state of $|1\ra$ and $|2\ra$. By symmetry, both states have the same energy (we set that to zero). If the quantum mechanical amplitude of tunneling is $-t'$, it is easy to observe that there are two stationary states in which the molecule can be present. One is the singlet state $\propto |1\ra-|2\ra$ with lower energy $-t'$, and the other is the triplet state $\propto|1\ra+|2\ra$ with higher energy $+t'$. This is exactly similar to what happens in the above case when C-type site forms virtual state and the resonance is over two O-type sites adjacent to it.

\subsection*{Case C: Loop resonance (figure 4C) and star resonance (figure 4D)}
In the loop resonance, the Hamiltonian in this case is given as: $H = 0|1\ra\la 1| + U |2\ra\la2| + 0 |3\ra\la3| + U |4\ra\la4| - t(|1\ra\la2| +  |2\ra\la3| +|3\ra\la4| + |4\ra\la1| +h.c.)$ which can be easily diagonalized to get the following set of the eigenvalues:

\begin{eqnarray}
\lambda_0 &=& 0\nonumber\\
\lambda_{1} &=& U\nonumber\\
\lambda_{-} &=& \frac{1}{2} (U - \sqrt{U^2 +16t^2})\nonumber\\
\lambda_{+}&=& \frac{1}{2} (U + \sqrt{U^2 +16t^2}).
\end{eqnarray}

Clearly, the lowest eigenvalue is $\lambda_{-}$ which in the said limit approximates to:

\beq
\lambda_{-} \simeq -4\frac{t^2}{U}.
\eeq

On exactly similar lines, star resonance (figure 4D) can be analyzed, and its lowest eigenvalue in the said approximation ($|t|<<U$) has the above stated value. Clearly then, it is energetically favorable for the doped particle to set into either star or loop resonance rather than to set into a "line" resonance of figure 4B.

\subsection*{Case D: Resonance over five states in a line: $|1\ra,~|2\ra,~|3\ra,~|4\ra, ~|5\ra$, where $|1\ra$ is on O-type cite, and $|5\ra$ is also on O-type state.}

The Hamiltonian in this case can be written as: $H = 0|1\ra\la 1| + U |2\ra\la2| + 0 |3\ra\la3| + U |4\ra\la4|  0 |5\ra\la5|- t(|1\ra\la2| +  |2\ra\la3| +|3\ra\la4| + |4\ra\la5| +h.c.)$. The smallest eigenvalue is:

\beq
\lambda_{-} = \frac{1}{2} (U - \sqrt{U^2 +12t^2}) \simeq -3 \frac{t^2}{U}
\eeq

It is very clear that if we have periodic boundary conditions (imagine that the sheet in figure 3 is folded in the form of a cylinder such that  alternate arrangement of C-type and O-type sites remains intact) and we consider the motion of the particle over a ring on the surface of the cylinder.  Suppose that its starts off from an O-type site and after full revolution it reaches to the same site and set in a resonance of this kind. It is very easy to see that resonance of this kind (ring resonance) will have the lowest eigenvalue (in the said approximation) given as:

\beq
\lambda_{-} \simeq - (n-2) \frac{t^2}{U},~~~n\ge3.
\eeq

Here $n$ is the total number of sites on the ring. Therefore we notice that a resonance over very extended sites leads to lower energy. So we conclude that rather than forming a local singlet (over two or three sites) the doped particle can exist in a very low energy eigenstates and these states are in the form of a resonance over a large number of extended sites (virtually over the entire lattice). Thus instead of forming a local singlet, the particle would like to "run away"! Or keep ``running" away!

\subsection*{Connection of this toy model with the real physics of $CuO_2$ planes}

The question is how does this simple toy model corresponds to the real physics of $CuO_2$ planes? The O-type sites are oxygen ions where we set $\ep_p =0$. We considered that C-type sites are copper ions with a single hole already existing there. When the doped hole migrates to the C-type site at energy $\ep_d$ it faces Hubbard repulsion $U$. The sum total of $\ep_d$ and $U$ we called an effective $U$. However, we have completely ignored the spin physics. But this is not correct if we consider the real physics of the $CuO_2$ planes. Spin fluctuations on the C-type lattice and spin exchange of the doped hole on O-type site with spins on C-type sites must be taken into account. If the characteristic frequency of the spin fluctuations on C-type sites is $\omega_{sf}$ and typical time scale of hole motion on the O-type sites is $\frac{\hbar}{t}$ ($t$ is the hoping amplitude), then one can consider three possible special cases: (1) fast spin-fluctuations case $(\omega_{sf} >>\frac{t}{\hbar})$; (2) intermediate case $(\omega_{sf} \sim \frac{t}{\hbar})$, and (3) slow spin fluctuations case $(\omega_{sf} <<\frac{t}{\hbar})$. The problem becomes much more interesting and complicated. In the fast fluctuations case $(\omega_{sf} >>\frac{t}{\hbar})$ the mobile hole will only "see" an average effect of spin fluctuations  (these are very fast on the timescale of hole's motion).  And the conclusions of the previous section (no local singlet formation) goes through. But for $\omega_{sf} \lesssim \frac{t}{\hbar}$, which is a realistic case pertaining to cuprates, it is difficult to make an a-priori guess. In addition, spin swaps between the spin of doped hole and that of the copper hole will take place.  The full analytic treatment is beyond the scope of this article (This is an interesting open problem which should be treated analytically). 

However, if we treat $U$ to be very large and consider states of the doped hole on C-type sites as virtual states (CASE B1 in the previous subsection), and further if the spin system of the C-type sites is "rigid" (no spin-fluctuations allowed), then the doped hole on O-type sites will "see" similar magnetic environment around it. Arguments of the CASE B1 in the previous subsection applies. {\it In this case, the doped hole will not settle in a local singlet but "but keep running away". } Now the question is:  how realistic is the "rigid magnetic lattice approximation"? It is shown by Mona Berciu and collaborators\cite{mona1}(will be discussed in section (VI)) that in the three-band model (Emery model in the strong correlation limit) if we treat the magnetic system as rigid (no spin fluctuations allowed) then the basic features of the hole dispersion can still  be reproduced. Thus, this case is relevant to cuprates, and the doped hole stays mobile. But, for the general case, we need to base our arguments on the experimental features.

\section{The two-component behaviour and the Barzykin-Pines analysis}

In this section we review Barzykin-Pines phenomenological analysis\cite{bp} which clearly points towards the three-band model (Zhang-Rice reduction seems incompatible with experimental features). It turns out that $CuO_2$ planes exhibit two-component behaviour. 

\subsection{What is the two-component behaviour?}

Based on various experimental signatures (such as the behaviour of static magnetic susceptibility: NMR Knight shift etc), Barzykin and Pines\cite{bp} argued that  the scaling behaviour of underdoped cuprates reflects the presence of two different components: (1) spin-liquid component, and (2) Fermi liquid like component. 

The scaling behaviour of the Spin Liquid (SL) component is that of a 2D Heisenberg model with quasi-localized spins on copper sites. Superexchange interaction between nearest neighbors spins is doping dependent and weakens as doping is increased:

\beq
J_{eff}(x) = J f(x),~~~J\sim1200~K.
\eeq
And $f(x) = 1-x/0.2$. This is intuitively easy to understand: the presence of doped holes on the oxygen sub-lattice weakens the spin-spin coupling of the copper ions. After the QCP around $p\simeq 0.2$ the spin liquid component disappears (in agreement with $p$ to $1+p$ transition\cite{tai}). This spin liquid component exists below a characteristic crossover temperature $T_{max}$. It is the temperature at witch the static magnetic susceptibility $\chi(T)$ exhibits a maximum when plotted as a function of temperature at a fixed doping $p<0.2$.

The other Fermi liquid (FL) like component grows in strength with increasing doping as $1-f(x) = x/0.2$. This FL like component exhibits temperature independent Pauli type susceptibility. Therefore the total susceptibility is written as:

\beq
\chi(x,T) = f(x) \chi_{SL}(T) + (1-f(x))\chi_{FL}.
\eeq

Here, $\chi_{SL}(T)$ is the static susceptibility of the spin liquid component and $\chi_{FL}(T)$ is the static susceptibility of the Fermi liquid component. This is the essence of the Barzykin-Pines two-component model.

\subsection{What is the physical interpretation of the two-component behaviour?}

As noted by V. J. Emery and others, doped holes go into oxygen $p$ orbitals and these constitute the mobile degrees-of-freedom. It forms the the Fermi Liquid like component (the second component) in the phenomenological description of $CuO_2$ planes by Barzykin and Pines. The presence of these mobile degrees-of-freedom weakens the superexchange coupling between the localized copper spins such that the effective exchange interaction reads $J_{eff}(x) = J f(x)$.  The localized copper spins constitute the spin liquid component. At finite doping ($p<0.2$) copper holes stay localized due to strong correlation effects (evidenced by $p$ to $1+p$ transition around $p^* \simeq 0.19$)\cite{tai}.  For $p>0.2$ the two-component behaviour disappears and have a large Fermi surface \cite{tai}.

These two components are not completely independent from each other. As V. J. Emery argues\cite{emery1}, due to hybridization, mobile $p$ holes pick up some $d$-character.

Barzykin and Pines have argued that the presence of $\chi_{SL}$ and $\chi_{FL}$ is incompatible with the single band Hubbard model and with the Zhang-Rice approximation of it (next section).

\subsection{Experimental signatures of the two-component behaviour}

\subsubsection{Static magnetic susceptibility and the Johnston-Nakano scaling}

At a given doping $p<0.2$ the static bulk spin susceptibility schematically looks like (figure 6).

\begin{figure}[h!]
    \centering
    \includegraphics[width=1.0\columnwidth]{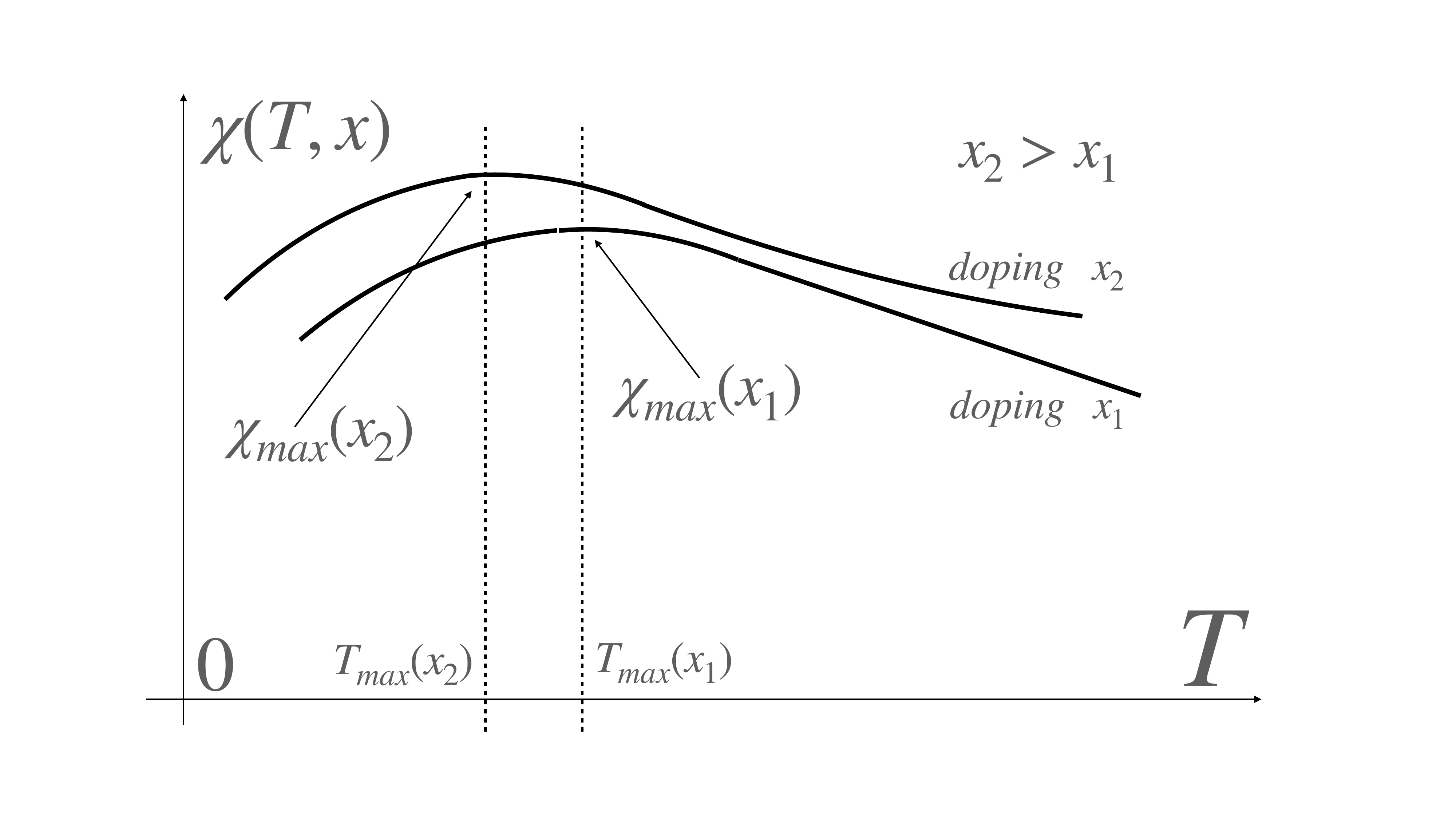}
    \caption{Schematic behaviour of the static magnetic susceptibility as a function of temperature in the underdoped regime.}
    \label{f6}
\end{figure}

Consider a doping value $x_1$. Starting from low temperatures it first increases and then exhibits a maximum $\chi_{max}(x_1)$ at a temperature $T_{max}(x_1)$, after which it drops with further increase in temperature. Similar behaviour is shown at a higher doping $x_2$, now $\chi_{max}(x_2)$ occurs at $T_{max}(x_2) < T_{max}(x_1)$.  For actual data refer to\cite{bp,john,nak}.

If the static susceptibility data is plotted on a scaled graph such that $\frac{\chi(T,x)}{\chi_{max}(x)}$ is plotted as function of $\frac{T}{T_{max}(x)}$ for various values of $x$, it turns out that {\it all the curves at various values of $x$ falls on a single universal scaling curve.} This scaling behaviour is called the Johnston-Nakano scaling behaviour.  It was first discovered by Johnston\cite{john} and refined and extended later by Nakano etal\cite{nak}. The static magnetic susceptibility can be represented by the following universal behaviour\cite{bp}: 

\beq
\chi(T,x) = \chi_0(x) + [\chi_{max}(x) - \chi_0(x)] \mathbb{F}\left(\frac{T}{T_{max}(x)}\right).
\eeq

The scaling function $\mathbb{F}(x)$ is a universal function and it matches with that calculated for 2D Heisenberg model in the spin liquid regime\cite{bp}. And $\chi_0(x)$ include the Fermi liquid and Van Vleck contributions.

The most important point to be noted is that the temperature dependent part $[\chi_{max}(x) - \chi_0(x)] \mathbb{F}\left(\frac{T}{T_{max}(x)}\right)$ which is originating from the spin liquid component disappears at $x\gtrsim0.2$. This is in agreement with the $p$ to $1+p$ transition around $p^* =0.19$\cite{tai}.  This clearly shows that the temperature dependent part of susceptibility is coming from the localized component which are copper spins interacting via renormalized exchange interaction $J_{eff} =J f(x)$. This is what Emery assumed (localized spins on copper sub-lattice and mobile holes on the oxygen sub-lattice)\cite{emery1}.

The Johnston-Nakano scaling clearly points towards the two-component behaviour, and thus the three-band Emery model where the interpretation is direct.  In this author's opinion, this observation experimentally established that the two-component behaviour which requires explicit consideration of a localized spin sub-system and a mobile sub-system which finds a natural justification in the three-band Emery model is not a "malum necessarium" (with respect to simple single-band and single-component description) but an essential minimal required component of the physics of high-Tc cuprates. Single band models leads to comparatively easier analysis but that is not the point. Single band models face a dismal failure in addressing the Barzykin-Pines phenomenology and other experimental features.

\subsubsection{The issue of cuprate NMR Knight shifts and the Mila-Rice-Shastry hyperfine scenario}

Another set of experimental observation which cannot be rationalized within the single band model is NMR Knight shift data\cite{bp, j1,j2,j3}. NMR resonance frequency of a nucleus depends on its spin, its gyromagnetic ratio, and applied magnetic field. Another important factor which changes (shifts) this resonance frequency is the electronic environment of the nucleus.  The effective magnetic field "seen" by the nucleus is equal to the sum of the external applied magnetic field and the induced (via local electronic distribution) magnetic field.  The induced magnetic field by the electronic environment is temperature dependent.  Thus, a shift in the NMR resonance frequency of a given nucleus can give important information regarding its magnetic environment coming from electrons. More precisely, the resonance frequency shift (the Knight shift) is proportional to the local electronic magnetic susceptibility:

\beq
\mck \propto \chi(T).
\eeq

Initial observations\cite{alloul1,taki} of Knight shifts at the nuclei $^{63}Cu,~~^{17}O,~~^{89}Y$ in $YBa_2Cu_3O_{6.63}$ showed very similar temperature dependences. From this, it appeared that a single magnetic susceptibility (originating from a single electronic fluid) is at work, such that

\beq
\mck_{\perp}^{63} \propto  \mck_{c}^{17} \propto\mck_{iso}^{17}  \propto\chi_{single}(T).
\eeq

From these observations originated the notion of the single-fluid scenario (single component scenario) which is consistent with the Zhang-Rice reduction and the single-band picture.  This was put on a rigorous foundation by the Mila-Rice-Shastry (MRS) hyperfine scenario\cite{mrs1,mrs2}. The single-fluid consists of quasi-localized copper spins and ZR-singlets mobile in a one-band reduction. 

The system $YBa_2Cu_4O_8$ also exhibits roughly similar temperature dependence for Knight shifts of the three nuclei $\mck_{\perp}^{63},~\mck_{c}^{17},~\mck_{iso}^{17},~\mck_{ax}^{17}$. The Knight shift studies on $YBa_2Cu_4O_8$ reinforced the notion of the single-component picture. 

However, the single-component picture faced the following problem: When the external applied magnetic field is parallel to the crystal c-axis (perpendicular to the ab-plane) the cooper Knight shift $\mck_{||}^{63}$ does not exhibit any temperature dependence, whereas $\mck_{\perp}^{63}$ (Knight shift when magnetic field is applied perpendicular to the c-axis or applied in the ab-plane) shows temperature dependent behaviour. How does one reconcile this behaviour within the single-component picture? 

This can be reconciled within the Mila-Rice-Shastry (MRS) hyperfine scenario\cite{mrs1,mrs2}. It turns out that there is an accidental cancellation between the direct and transferred hyperfine terms in YBCO systems. For copper nucleus one can write:

\begin{eqnarray}
\mck_{||}^{63}(T) &=& [A_{||} +4 B] \chi_{single}(T)] + const.\nonumber\\
\mck_{\perp}^{63}(T) &=& [A_{\perp} +4 B] \chi_{single}(T)] + const.
\end{eqnarray}

There is a single temperature dependent susceptibility $\chi_{single}(T)$ which couples to copper nuclei via two terms. $A_{||}$ is the direct (on site) hyperfine coefficient and $B$ is the transferred one from oxygen orbitals to the copper nuclei. It is proposed that

\begin{eqnarray}
A_{||} +4 B = 0\nonumber\\
A_{\perp} +4 B \ne 0.
\end{eqnarray}

That is, an accidental cancellation removes the temperature dependence of $\mck_{||}^{63}$, but $\mck_{\perp}^{63}(T)$ retains the same temperature dependence. Thus, within this Mila-Rice-Shastry transferred hyperfine scenario, the single-component picture was retained.

\subsubsection{The failure of the single-component model}

The first blow to the single-component scenario came from the Knight shift observations in $La_{2-x}Sr_xCuO_4 ~(LSCO)$\cite{j2} (and references therein). The temperature dependence of the Knight shift for $^{63}Cu$ nucleus is very different from that of $^{17}O$ nucleus. It seems that two different fluid components are active in the $CuO_2$ planes of $LSCO$ (one fluid at copper sites and other at oxygen sites). In other words, both shifts are not proportional to each other.

It turns out that the susceptibility that dominates the copper shifts is almost independent of temperature (Fermi liquid like behaviour) and it falls off very quickly near $T_c$. The other susceptibility (that dominates the oxygen shifts) is temperature dependent. These observations cannot be understood within a single-component scenario. Also, it is to be noted that the behaviour of LSCO is very anomalous\cite{j2}. In the Barzykin-Pines model the susceptibility that is active at copper sites (originating from the spin liquid component) is temperature dependent whereas  the susceptibility coming form oxygen holes is temperature independent  (Fermi liquid like component).  What is seen in the LSCO is just the opposite. This remains an important open problem in the field.

In addition, the system $HgBa_2CuO_{4+\delta}$ also clearly violates the hyperfine scenario\cite{j2}. In this case even $A_{||} +4 B \ne 0$. That is, the proposed accidental cancellation is missing.

NMR relaxation studies also point towards the failure of the one-component scenario\cite{j4}. However, phenomenology of the NMR relaxation times is much more complex.  Interested readers can refer to\cite{bp,j4}. The take-home message from the Knight shift phenomenology is that the single-band models (including the ZR reduction) fail to account for these experimental features.

\section{Single hole motion in one-band and three band models.}

Mona Berciu and collaborators have studied the charge dynamics of a single doped hole both in one-band and three-band models using a variational approximation that include spin fluctuations in the vicinity of the doped hole\cite{mona1,mona2}. For the single band case they studied $t-t'-t^{\p\p}-J$ model. It is well known that $t-J$ model fails to reproduce the quasi-particle (qp) dispersion in the $(0,\pi) -- (\pi,0)$ direction in the Brillouin  zone (there is a sizable curvature in the dispersion along this direction as seen through ARPES experiments whereas the $t-J$ model predicts a nearly flat dispersion, in contradiction to experiments). However, in  $t-t'-t^{\p\p}-J$ model due to longer range (next nearest and next-to-next neighbor) hopings, the qp dispersion is faithfully reproduced. In fact, Mona Berciu and collaborators noticed that both the spin-fluctuations (near the doped hole) and the longer range hopping are required to address the  complete qp dispersion (i.e., these have complementary contributions).

But, for the three-band Emery model in the strong correlation limit (within the same variational approximation) authors\cite{mona1,mona2} noticed that spin fluctuations (in the vicinity of the doped hole) has a minor effect on the qp dispersion and on the qp particle dynamics. The authors\cite{mona1,mona2} conclude that these two models (one-band versus three bands) exhibit very different qp dynamics (even when the qp dispersion is matched), and the reduction of the three-band model to one-band model is questionable. 

Authors further argue that the hoped for simplification (from the Emery model to $t-t^\p-t^{\p\p}-J$ model) leads to different qp dynamics of the doped hole in these two models, even though both can reproduce the qp dispersion.  In the vantage point of the NMR Knight shift phenomenology and other inputs, we can argue that three-band Emery model predicts the likely correct qp dynamics whereas the single-band model fails.

Technical details of the analysis\cite{mona1,mona2} can be briefed in the following way: (a) consider the case of $t-t'-t^{\p\p}-J$ model with fixed magnetic background (no spin fluctuations allowed). The qp dispersion obtained via the variational method\cite{mona1,mona2} can be schematically shown (figure 7 A). It is found that with the fixed magnetic background, $t-t'-t^{\p\p}-J$ model fails to capture the dispersion between $(0,0)$ to $(\pi,\pi)$ direction, that is, along the nodal line of the Brillouin zone. 
\begin{figure}[h!]
    \centering
    \includegraphics[width=1.0\columnwidth]{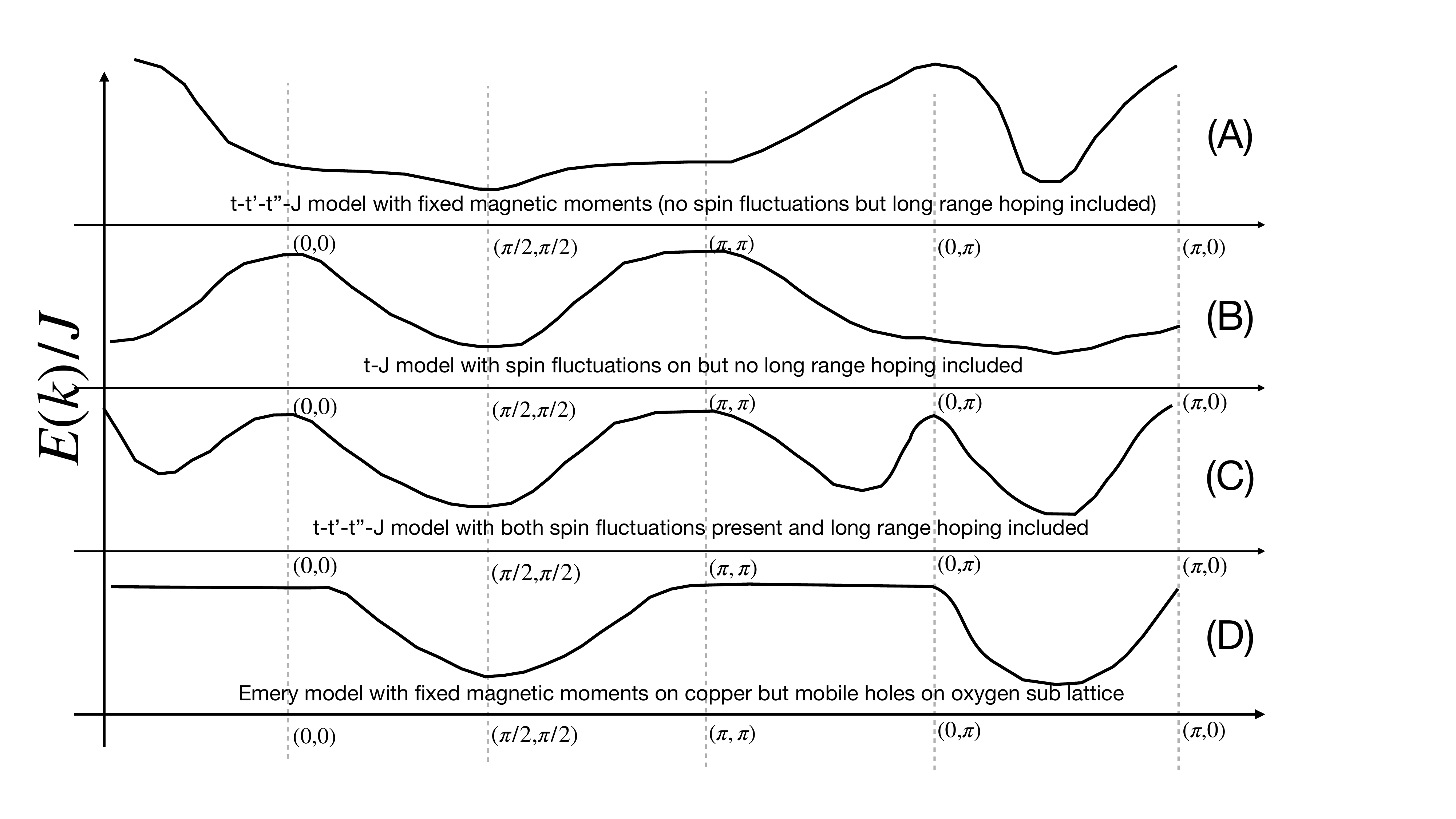}
    \caption{Schematic behaviour of the quasiparticle dispersion in various sub-cases.}
    \label{f8}
\end{figure}
If, on the other hand, no longer range hoping is allowed (that is, the bare $t-J$ model is used) but spin fluctuations are turned on near the doped hole, then the model fails to capture the dispersion along the ($0,\pi$) to $(\pi,0)$ direction (figure 7B). Thus leaving one of the aspect (either longer range hoping or spin fluctuations) leads to inadequate reproduction of the qp dispersion in the t-J model. Spin-fluctuations are introduced by flipping the spins of the nearest-neighbor AFM pairs through the Hamiltonian $H_{sf} = \frac{J}{2} \sum_{<i,j>} (S^{-}_i S_j^{+} + S_i^{+} S_j^{-})$, where $S_i^{+}$ is a spin at site $i$ pointing in the $+z$-direction. Flipping a spin in the AFM background leads to pairs of spins in the same direction: ferromagnetic pairs (these are called  "still magnons"). It is found that these "still magnons" are essential in the $t-J$ model to get the correct dispersion along the nodal line $(0,0)--(\pi,\pi)$.

When both the spin fluctuations and longer range hopings are used in the $t-J$ model (that is, $t-t'-t^{\p\p}-J$ with spin fluctuations included), the qp dispersion obtained agrees very well with the experimental results (figure 7C). The said complementarity\cite{mona1,mona2} in the roles of the spin fluctuations and the longer range hoping in the $t-J$ model is that spin-fluctuations address the part of the dispersion along the  $(0,0)--(\pi,\pi)$ direction whereas longer range hopings address the part of the dispersion along the ($0,\pi$) to $(\pi,0)$ direction.

By contrast, in the three-band Emery model, even when the spins on the copper sub-lattice are frozen (no flip fluctuations allowed) the qp particle dispersion is well reproduced both along the $(0,0)--(\pi,\pi)$ direction and along the ($0,\pi$) to $(\pi,0)$ direction (figure 7D). Thus authors conclude that the hoped for simplification (from the Emery model to $t-t^\p-t^{\p\p}-J$ model) leads to very different qp dynamics (especially related to spin-fluctuations and longer range hoping) of the doped hole in these two models, even though both can reproduce the qp dispersion.  {\it This is a very important point and raises questions on the existence of the Zhang-Rice singlets. Including spin fluctuations in the three-band Emery model reproduces the dispersion very well. But important point to be noted is that even without spin fluctuations in the Emery model, the essential features of the dispersion (both along $(0,0)--(\pi,\pi)$  and along$(0,\pi)$ to $(\pi,0)$) are already reproduced.} Thus, spin fluctuations are less relevant in the three band model (as discussed in section IV). The authors\cite{mona1} summarize the whole scenario it in perfect way: 

\begin{quote}
{\texttt{``...obtaining the correct dispersion for quasiparticles is not sufficient to validate that model. The dispersion can have correct shape for the wrong reasons."-----Authors\cite{mona1} } }
\end{quote}

{\it Therefore, the reduction from the three-band Emery model to $t-t'-t^{\p\p}-J$ model is not justified. }

\section{Connection between the Gor'kov-Teitel'baum phenomenological model and the three-band Emery model}

Gor'kov and Teitel'baum rationalized the temperature dependence of the Hall coefficient of underdoped cuprates using a simple-minded phenomenological model (now called the Gor'kov-Teitel'baum Thermal Activation (GTTA) model). They argued that Hall coefficient can be expressed as:

\beq
R_H = \frac{1}{n_{Hall} e c}.
\eeq
Here, $n_{Hall}$ is the effective carrier density which is both doping and temperature dependent:

\beq
n_{Hall}(x,T) = n_0(x) +n_1 e^{-\frac{\Delta(x)}{T}}.
\eeq

It has two components: (1) doping dependent but temperature independent term $n_0(x)$, and (2) both doping and temperature dependent term $n_1 e^{-\frac{\Delta(x)}{T}}$. The temperature independent term $n_0(x)$ is linearly proportional to doping value $x$ at low doping range (for LSCO, it is approximately linear upto $x=0.1$) and for higher doping it is super-linear. The coefficient $n_1$ in the second term is almost temperature independent and drops to zero near the QCP ($p^*\simeq 0.19$). The temperature dependent term $n_1 e^{-\frac{\Delta(x)}{T}}$ represents a thermal activation character. According to Gor'kov and Teitel'baum it represents thermal excitation of quasiparticles from van Hove points $(0,\pm \pi)$ to the nodal arcs ($\pm \pi,\pm,\pi$) at the chemical potential. The activated quasiparticles then join already existing (from doped holes) mobile carriers.

The validity of the GTTA model is checked through the deduction of the doping dependent gap $\Delta(x)$ purely from the Hall coefficient data and its doping dependence agrees very well with that of the pseudogap obtained from a completely different experiment (ARPES). Thus $\Delta_{GTTA}(x) = \Delta_{PG}(x)$ (for LSCO, refer to\cite{gtta}; and for other cuprates, refer to\cite{nav1}). For a brief review of the GTTA model refer to\cite{nav2}.

Lev Gor'kov\cite{gor} argued that doped holes go to the Fermi arcs (the length of the arc is roughly proportional to the doping value $x$ at a fixed temperature). These mobile holes on the arcs constitute Fermi liquid behaviour. Now, if we recall Emery's point of view (that is, mobile carriers reside on the oxygen sub-lattice), then we can understand that these doped holes reside on the oxygen sub-lattice (in the real space).  In $k-$space these constitute "Fermi arcs". In fact Gor'kov explicitly shows that the resistivity due to these mobile holes on Fermi arcs is given by $\rho \sim T^2$ for $T<T^{**}<T^*$. Here, $T^{**}$ is another temperature scale inside the PG regime. The scaling $\rho \sim T^2$ has actually been observed experimentally\cite{navin}.

For the calculation, Gor'kov simplifies the qp dispersion for holes on the Fermi arcs by $\ep(p) \simeq v_F (p-p_F)$ and used the Boltzmann kinetic equation for electron-electron scattering to calculate the resistivity. Explicit inclusion of U-processes lead to $T^2$ resistivity. On the other hand, if one tries to calculate resistivity due to electron-electron scattering with U-processes taken into account for momentum degradation in a single band (say in $t-t'-t^{\p\p}-J$) model, then it is clear that spin-fluctuation effects cannot be neglected (these are required to address the dispersion along the $(0,0)--(\pi,\pi)$ direction\cite{mona1}). Therefore, one needs to show in a single band model with the consideration of spin-fluctuations with or without U-processes that $\rho\sim T^2$. As far as this author knows, this problem has not been addressed within the above mentioned constraints.

On the other hand, in three band Emery model, with magnetic lattice frozen in the low doping regime (reasonably good approximation to address qp dispersion), holes on oxygen sub-lattice can undergo Fermion-Fermion scattering with U-processes and can give $T^2$ resistivity (as shown by Gor'kov\cite{gor}).  At the same time, at higher doping the "magnetic lattice" is no more rigid as the effective exchange interaction is reduced $Jf(x)$ ($\grave{a} ~la~ $  Barzykin-Pines). At optimal doing, the fluctuating moments on copper atoms seems to play a crucial role in the dead $T-$linear resistivity seen at that doping. These issues remain a difficult open problem in the field\cite{ns1,ns2}.  The idea of two-electronic subsystems is also worked out based on transport data in\cite{sunko}.

However, in the underdoped case, two lessens can be learned from the GTTA model:

\begin{enumerate}
\item Temperature dependent part of the effective carrier density is coming from localized copper spins. This term disappears at $p = p^*\simeq 0.19$($n_1$ drops abruptly at this critical doping) in agreement with $p$ to $1+p$ transition at $p^*$\cite{tai}.
\item Mobile holes reside on the Fermi arcs in $k-$space and on oxygen sub-lattice in real space. 
\end{enumerate}

Thus, the Emery's division of the total electronic system of $CuO_2$ planes into the mobile degrees-of-freedom and the localized spin degrees-of-freedom is compatible with the Gor'kov-Teitel'baum phenomenological model.

\section{Unconventional superconductivity in one-band and three band models: a brief summary of recent numerical studies}

It is observed that $t-J$ model and one-band Hubbard model exhibit many ``near degenerate" ground states\cite{tj1,tj2,tj3,tj4,tj5,tj6,tj7,tj8,tj9,tj10,tj11,tj12,tj13,tj14,tj15,tj16,tj17,tj18,tj19,tj20}. The d-wave superconductivity completes with other competing orders such as stripes and density waves. In addition, $t-J$ model cannot reproduce the quasiparticle dispersion\cite{mona1,mona2}. As seen, incorporation of the longer range hopping terms into the $t-J$ model can rectify the problem of the quasiparticle dispersion\cite{mona1,mona2}. The question is regarding its superconducting instability.  In an important investigation using DMRG\cite{jia}, it is seen that the extended $t-J$ model (that is, $t-t'-t^{\p\p}-J$ model) exhibits suppression of pairing for ($t'<0,t^{\p\p}>0$) which is associated with hole doping, whereas for  ($t'>0,t^{\p\p}<0$) which is associated with electron doping it shows stronger pairing! But, this is in direct contradiction to what is observed in experiments (hole doping side shows stronger superconductivity). Therefore, it is argued\cite{jia} that the reduction of the three-band model via ZR singlet picture to the $t-J$ model (or to $t-t'-t^{\p\p}-J$ model) misses the central point. In authors\cite{jia} words:

\begin{quote}
{\texttt{``These results indicate a fundamental flaw in the model [ $t-t'-t^{\p\p}-J$ model] for reliable description of the doping dependence of pairing in the Cuprates"--- Authors\cite{jia}.
}}
\end{quote}

But, in the case of three-band Emery model, using determinant quantum monte carlo (DQMC) and dynamical cluster approximation (DCA),  authors\cite{mai} show that superconducting dome can be reproduced both on the hole-doping side and on the electron-doping side. Authors\cite{mai} also find an optimal value of the charge transfer gap ($\Delta_{CT}$) at which $T_c$ exhibits a maximum. A more physical understanding of why this happens is lacking.

On the other hand, it is shown that one-band Hubbard model with next nearest neighbor hopping ($t-t^\p-U$ not its descendent $t-t^\p-J$ model) exhibits dome shaped order-parameter behaviour on both the sides of the phase diagram\cite{xu}. Authors\cite{xu} use state of the art constrained path (CP) auxiliary field quantum monte carlo (AFQMC) and DMRG methods. It is observed that on the hole doped side superconductivity is stronger and co-exists with fractionally filled stripe correlations. On the electron doped side, a weakly modulating AFM phase coexists with a weaker form of superconductivity. Thus, this study\cite{xu} shows that the parent Hubbard model performs better than its  descendents. The question is:  how does the parent $t-t^\p-U$ model perform  with respect to quasiparticle dispersion and dynamics\cite{mona1}.

Very recently, Authors\cite{chen} argue that $t-J$ model is a proper minimal model to address superconductivity in cuprates. Using DMRG calculations on large 8-leg cylinders, they show that $t-J$ model itself is sufficient to explain robust d-wave superconductivity possibly coexisting with weak pair density wave order. Authors\cite{chen} argue that pair-density and charge density waves are suppressed with a small next nearest neighbor hopping $t^\p$ and d-wave superconductivity is uniform. But again, it is clear that $t-J$ model is a dismal failure in addressing the experimental phenomenology (in the previous sections) and it also fails to reproduce the correct qp dispersion\cite{mona1}.

Therefore, we notice that there are conflicting results in the literature as far as numerical studies of the $t-J$ model  and other one-band models are concerned. As questioned in ref\cite{xu}: whether there is any analytic theory of cuprate superconductivity which can help resolve all these issues? Or simulation is the only way to proceed? 

{\it In this authors opinion, only those models (such as Emery model) which are able to address the detailed experimental phenomenology (Barzykin-Pines and Gor'kov-Teitel'baum etc) should be put to computers and simulations. The analytical theory of cuprate superconductivity can be found once strict scrutiny of the models in the light of experimental phenomenology is done. }

\section{Conclusion} 

The author has reviewed an overwhelming amount of evidence which do not support Zhang-Rice singlet formation, and the single-band models face a dismal failure in addressing the essential experimental phenomenology (that is, the NMR Knight shifts, the Barzykin-Pines phenomenology, the Gor'kov-Teitel'baum phenomenology etc). A simple toy model, introduced by the author, highlights that rather than forming local singlets, doped holes would tend to settle in extended resonances or simply be mobile in the $CuO_2$ lattice. This brings us to the original insight of V. J. Emery and G. Reiter:

\begin{quote}
{\texttt{``....they claim that the extended Hubbard model [the Emery model]  we have been considering is "equivalent" to the one-band Hubbard model, the singlet [ZR singlet] being equivalent to a vacancy. This is not correct."------Emery and Reiter\cite{final}.
}}
\end{quote}

\begin{acknowledgments}

Author is thankful to Udit Khanna for carefully reading the draft and providing comments. Through this article, author also pays his tribute to J. V. Narlikar (19 July 1938 -- 20 May 2025) who passed away recently. Prof. Narlikar has done great work in science popularization in India in addition to his fundamental work in cosmology.

\end{acknowledgments}



\begin{thebibliography}{99}
\bibitem{comment1} Anderson later disowned his book when he realized that his inter-layer theory was not correct\cite{andrew}. But he maintained that the one-band model of cuprates is able to capture the basic mechanism of superconductivity in cuprates\cite{ander1}. Still the book has many novel observations.
\end{thebibliography}
\end{document}